# Single-state multiparty semiquantum secret sharing with *d*-dimensional Bell states

Ying Chen, Zhi-Gang Gan, Tian-Yu Ye*
College of Information & Electronic Engineering, Zhejiang Gongshang University, Hangzhou 310018, P.R.China
E-mail：yetianyu@zjgsu.edu.cn (T.Y.Ye)

**Abstract:** A single-state multiparty semiquantum secret sharing (MSQSS) scheme with *d*-dimensional Bell states is proposed, which can accomplish the goal that only when all receivers work together can they restore the sender's secret key. This protocol is validated to be secure against both the outside attack and the participant attack. This protocol is adaptive for the *d*-dimensional system, only employs one kind of *d*-dimensional Bell states as initial quantum resource and needs neither quantum entanglement swapping nor unitary operations.

## 1 Introduction

With the expanding practical application of quantum information science, quantum cryptography has been developed rapidly. Over the last several decades, scholars have been constantly researching on different branches of quantum cryptography, such as quantum key distribution (QKD) [1-2], quantum secure direct communication (QSDC) [3-4], quantum dialogue [5-10], quantum key agreement (QKA) [11-12], quantum secret sharing (QSS) [13-20], *etc*. Here, QSS, first put forward by Hillery *et al.* [13] in 1999, is the generalization of classical secret sharing into quantum scenario, whose goal is that only when all receivers work together can they restore the sender's secret key.

Beside quantum cryptography, semiquantum cryptography has arisen during recent years, which allows partial users to only possess limited quantum capabilities. In 2007, Boyer *et al.* [21] put forward the 'semiquantumness' concept for the first time in a novel semiquantum key distribution (SQKD) protocol with single photons. Later, single photons in two degrees of freedom have also been used to design SQKD protocols [22,23]. A restricted classical user is only permitted to perform the following four operations [21]:(a) to send particles via the quantum channel; (b) to measure particles in the $Z$ basis (i.e., $\{|0\rangle,|1\rangle\}$ ); (c) to prepare particles in the $Z$ basis ; and (d) to scramble particles.

Semiquantum secret sharing (SQSS), as the combination of the 'semiquantumness' concept and QSS, have been developed greatly during recent years. Various SQSS protocols [24-32] have been proposed from different perspectives. Unfortunately, the overwhelming majority of previous SQSS protocols [24-31] cannot be applicable to $d$ -dimensional quantum system, where $d > 2$ . In other words, the SQSS protocols of Ref.[32] are the only two SQSS schemes feasible for *d*-dimensional quantum system.

Based on the above analysis, in this paper, we aim to construct a single-state multiparty semiquantum secret sharing (MSQSS) protocol by using $d$ -dimensional Bell states. The proposed protocol is applicable to the *d*-dimensional quantum system, only employs one kind of *d*-dimensional Bell states as initial quantum resource, and employs neither quantum entanglement swapping nor unitary operations.

## 2 Protocol description

The $d$-dimensional Bell states can be expressed as

$$\left|\psi_{u,v}\right\rangle=\frac{1}{\sqrt{d}}\sum_{k=0}^{d-1}e^{\frac{2\pi iku}{d}}\left|k\right\rangle\left|k\oplus v\right\rangle, \tag{1}$$

where $u,v\in\{0,1,\ldots,d-1\}$ and the symbol $\oplus$ represents the addition modulo $d$. Apparently, it has

$$\left|\psi_{00}\right\rangle=\frac{1}{\sqrt{d}}\sum_{k=0}^{d-1}\left|k\right\rangle\otimes\left|k\right\rangle=\frac{\left|0\right\rangle\left|0\right\rangle+\left|1\right\rangle\left|1\right\rangle+\ldots+\left|d-1\right\rangle\left|d-1\right\rangle}{\sqrt{d}}. \tag{2}$$

In the $d$-dimensional quantum system, the $Z$ basis and the $X$ basis can be depicted as

$$Z=\{\left|0\right\rangle,\left|1\right\rangle,\ldots,\left|d-1\right\rangle\} \tag{3}$$

and

$$X=\{F\left|0\right\rangle,F\left|1\right\rangle,\ldots,F\left|d-1\right\rangle\}, \tag{4}$$

respectively. Here, $F$ represents the discrete quantum Fourier transform, where

$$F\left|j\right\rangle=\frac{1}{\sqrt{d}}\sum_{k=0}^{d-1}e^{\frac{2\pi ijk}{d}}\left|k\right\rangle \tag{5}$$

for $j=0,1,\ldots,d-1$. Apparently, the $Z$ basis and the $X$ basis are two common conjugate bases.

Suppose that $P_0$ is the party with full quantum capabilities, while $P_1,P_2,\ldots,P_N$ are $N$ parties only possessing limited quantum capabilities. The goal of the proposed single-state MSQSS protocol with $d$-dimensional Bell states is that only when $P_1,P_2,\ldots,P_N$ collaborate together can they reveal $P_0$ 's secret key. The proposed single-state MSQSS protocol can be described as follows.

Step 1: $P_0$ prepares $N$ $d$-dimensional Bell state sequences of length $4n$, where each Bell state is in the state $\left|\psi_{00}\right\rangle$. Let $S_i^j$ and $T_i^j$ represent the first and the second particles of the $j$-th $d$-dimensional Bell state in the $i$ th sequence, respectively, where $i=1,2,\ldots,N$ and $j=1,2,\ldots,4n$. Note that according to Eq.(2), the first particle and the second particles of each $d$-dimensional Bell state are in the same state. Moreover, let $S_i=\{S_i^1,S_i^2,\ldots,S_i^{4n}\}$ and $T_i=\{T_i^1,T_i^2,\ldots,T_i^{4n}\}$, where $i=1,2,\ldots,N$. Then, $P_0$ sends the particles of $S_i$ to $P_i$ one by one via the quantum channel, and keeps $T_i$ in her own hand. Note that except the first particle of $S_i$, $P_0$ sends out the next one only when she receives the previous one.

Step 2: Upon receiving each particle of $S_i$, $P_i$ randomly chooses either to measure it with the $Z$ basis, prepare a new one in the found state and send it to $P_0$ (referred as MEASURE), or to reflect it back to $P_0$ without disturbance (referred as REFLECT). Here, $i=1,2,\ldots,N$.

Step 3: $P_0$ temporarily stores all of the particles from $P_i$, where $i=1,2,\ldots,N$. Then, $P_i$ announces for which particles in $S_i$ she chose to MEASURE. Two different Cases should be described.

Case (1): with respect to the particles in $S_i$ for which $P_i$ chose to MEASURE, $P_0$ measures the particles in $S_i$ and the corresponding particles in $T_i$ on her site with the $Z$ basis. The number of MEASURE particles is $2n$ . For the sake of security check, $P_0$ randomly selects half of MEASURE particles in $S_i$ and tells $P_i$ their positions. After that, $P_i$ informs $P_0$ of her measurement results on these chosen MEASURE particles. Afterwards, $P_0$ judges whether the quantum channel is secure or not by comparing her measurement results on these chosen MEASURE particles in $S_i$ , her measurement results on the corresponding particles in $T_i$ , and $P_i$ 's measurement results on these chosen MEASURE particles in $S_i$ . When no eavesdropper exists in the quantum channel, these measurement results should always be correspondingly identical.

Case (2): with respect to the particles in $S_i$ for which $P_i$ chose to REFLECT, $P_0$ judges whether the quantum channel is secure or not by performing *d*-dimensional Bell basis measurement on these particles and the corresponding particles in $T_i$ . When no eavesdropper exists in the quantum channel, $P_0$ 's Bell basis measurement results should always be $|\psi_{00}\rangle$ .

Step 4: $P_0$ checks the error rates in Cases (1) and (2). If the error rate in any Case is abnormally high, the protocol will be terminated immediately; otherwise, it will be continued.

Step 5: The left $n$ MEASURE particles in $S_i$ are utilized to share secret, where $i = 1,2,\ldots,N$ . Let $K_i$ denote the classical values corresponding to $P_0$ 's measurement results on the left $n$ MEASURE particles in $S_i$ , where $K_i \in \{0,1,\ldots,d-1\}$ and $i = 1,2,\ldots,N$ . Note that $|0\rangle,|1\rangle,\ldots,|d-1\rangle$ are encoded into the classical values $0,1,\ldots,d-1$ , respectively, here. As a result, $P_0$ makes $K = K_1 \oplus K_2 \oplus \ldots \oplus K_N$ be her secret key. Apparently, $P_i$ can automatically get $K_i$ . Hence, only when $P_1,P_2,\ldots,P_N$ collaborate together can they recover $K$ .

## 3 Security analysis

### 3.1 Outside attack

(1) Entangle-measure attack

In order to obtain $K$ , an outside eavesdropper, Eve, should know $K_i$ first, where $i = 1,2,\ldots,N$ . Eve may try her best to get $K_i$ through the entangle-measure attack depicted in Fig.1, which includes two unitaries, $U_E$ and $U_F$ . Here, Eve exerts $U_E$ on the particles sent from $P_0$ to $P_i$ and $U_F$ on the particles returned from $P_i$ to $P_0$ . Moreover, $U_E$ and $U_F$ share a common probe space having the initial state $|\varepsilon\rangle$ . As illustrated in Ref.[21], the shared probe allows Eve to perform the attack on the returned particles from $P_i$ to $P_0$ with the help of the information gained from $U_E$ ; and any attack where $U_F$ depends on a measurement after $U_E$ can be accomplished by $U_E$ and $U_F$ with controlled gates.

***Theorem 1:*** Suppose that Eve performs $(U_E, U_F)$ on the particles from $P_0$ to $P_i$ ( $i = 1,2,\ldots,N$ ) and from $P_i$ to $P_0$ . To introduce no error in Step 3, the final state of Eve's probe should be irrelevant to not only the operation of $P_i$ but also the measurement results of both $P_0$ and $P_i$. In this way, Eve will know nothing about $K_i$ by launching this attack if not being detected.

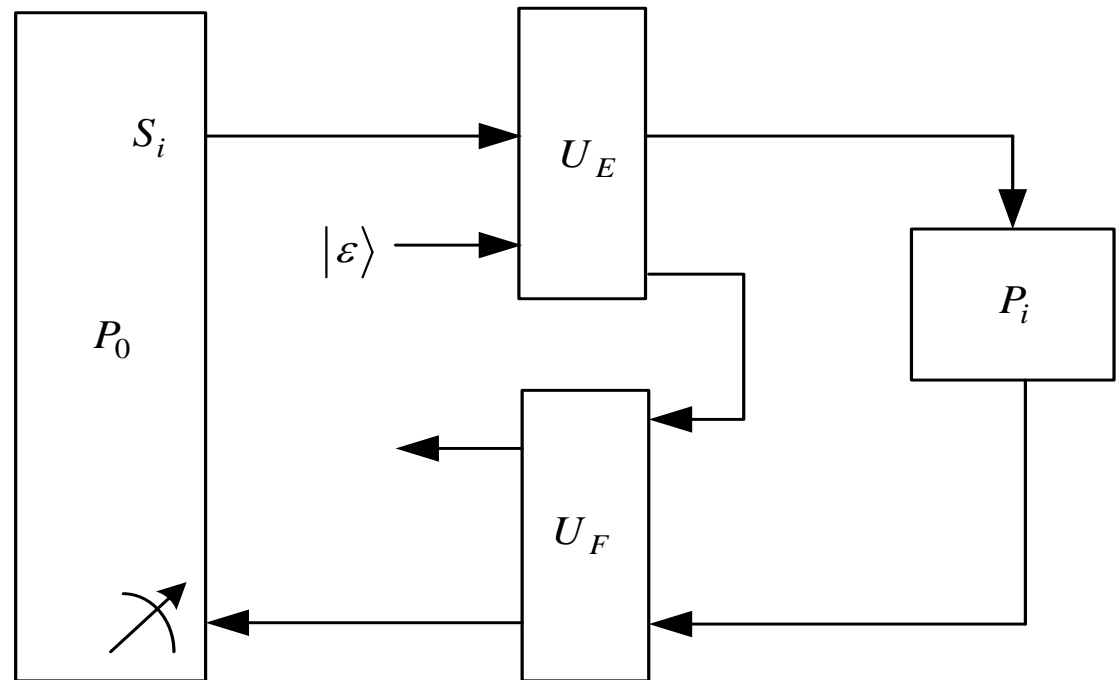

Fig.1 Eve's entangle-measure attack with two unitaries, $U_E$ and $U_F$

***Proof:*** The effect of $U_E$ on the particle $|k\rangle$ can be described as

$$U_E\left(|k\rangle|\varepsilon\rangle\right)=\sum_{t=0}^{d-1}\beta_{kt}|t\rangle|\varepsilon_{kt}\rangle\ . \tag{6}$$

Here, $|\beta_{kt}\rangle$ is Eve's probe state dependent on $U_E$, where $k,t=0,1,\ldots,d-1$. Moreover, for $k=0,1,\ldots,d-1$, $\sum_{t=0}^{d-1}|\beta_{kt}|^2=1$.

Before Eve's attack, the global state of the composite system composed by the particles of $P_0$ and Eve can be described as $|\psi_{00}\rangle|\varepsilon\rangle$. After Eve performs $U_E$, the composite system is evolved into

$$U_E\left(|\psi_{00}\rangle|\varepsilon\rangle\right)=\frac{1}{\sqrt{d}}\sum_{k=0}^{d-1}U_E\left(|k\rangle|\varepsilon\rangle\right)|k\rangle=\frac{1}{\sqrt{d}}\sum_{k=0}^{d-1}\left(\sum_{t=0}^{d-1}\beta_{kt}|t\rangle|\varepsilon_{kt}\rangle\right)|k\rangle=\frac{1}{\sqrt{d}}\sum_{t=0}^{d-1}|t\rangle\left(\sum_{k=0}^{d-1}\beta_{kt}|k\rangle|\varepsilon_{kt}\rangle\right). \tag{7}$$

When $P_i$ gets the particle of $S_i$, she chooses either to MEASURE or to REFLECT.

① Consider the situation that $P_i$ has chosen to MEASURE. According to Eq.(7), when $P_i$'s measurement result on the particle of $S_i$ is $|t\rangle$, the state of the composite system is collapsed into $|t\rangle\left(\sum_{k=0}^{d-1}\beta_{kt}|k\rangle|\varepsilon_{kt}\rangle\right)$, where $k,t=0,1,\ldots,d-1$.

Eve performs $U_F$ on the particle from $P_i$ to $P_0$. For Eve not being detectable in Case (1) of Step 3, $U_F$ should satisfy

$$U_F\left[|t\rangle\left(\sum_{k=0}^{d-1}\beta_{kt}|k\rangle|\varepsilon_{kt}\rangle\right)\right]=|t\rangle|t\rangle|F_t\rangle\ , \tag{8}$$

which means that $U_F$ cannot change the states of the particle in $S_i$ from $P_i$ to $P_0$ after $P_i$'s operation and the corresponding particle in $T_i$ on $P_0$'s site.

② Consider the situation that $P_i$ has chosen to REFLECT. In this situation, according to Eq.(7), the state of the composite system is $\frac{1}{\sqrt{d}}\sum_{t=0}^{d-1}|t\rangle\left(\sum_{k=0}^{d-1}\beta_{kt}|k\rangle|\varepsilon_{kt}\rangle\right)$, where $k,t=0,1,\ldots,d-1$.

Eve performs $U_F$ on the particle from $P_i$ to $P_0$. The state of the composite system is turned into

$$U_F\left[U_E\left(|\psi_{00}\rangle|\varepsilon\rangle\right)\right]=U_F\left[\frac{1}{\sqrt{d}}\sum_{t=0}^{d-1}|t\rangle\left(\sum_{k=0}^{d-1}\beta_{kt}|k\rangle|\varepsilon_{kt}\rangle\right)\right]=\frac{1}{\sqrt{d}}\sum_{t=0}^{d-1}U_F\left[|t\rangle\left(\sum_{k=0}^{d-1}\beta_{kt}|k\rangle|\varepsilon_{kt}\rangle\right)\right]. \quad (9)$$

Applying Eq.(8) into Eq.(9) generates

$$U_F\left[U_E\left(|\psi_{00}\rangle|\varepsilon\rangle\right)\right]=\frac{1}{\sqrt{d}}\sum_{t=0}^{d-1}|t\rangle|t\rangle|F_t\rangle. \quad (10)$$

For Eve not being detectable in Case (2) of Step 3, the probability that $P_0$ 's measurement result is $|\psi_{00}\rangle$ should be 1. Thus, by virtue of Eq.(2) and Eq.(10), we can get

$$|F_0\rangle=|F_1\rangle=\ldots=|F_{d-1}\rangle=|F\rangle. \quad (11)$$

③ Applying Eq.(11) into Eq.(8) produces

$$U_F\left[|t\rangle\left(\sum_{k=0}^{d-1}\beta_{kt}|k\rangle|\varepsilon_{kt}\rangle\right)\right]=|t\rangle|t\rangle|F\rangle. \quad (12)$$

By virtue of Eq.(2), applying Eq.(11) into Eq.(10) creates

$$U_F\left[U_E\left(|\psi_{00}\rangle|\varepsilon\rangle\right)\right]=\frac{1}{\sqrt{d}}\sum_{t=0}^{d-1}|t\rangle|t\rangle|F\rangle=|\psi_{00}\rangle|F\rangle. \quad (13)$$

According to Eq.(13) and Eq.(14), it can be concluded that for inducing no error in Step 3 , the final state of Eve's probe should be irrelevant to not only the operation of $P_i$ but also the measurement results of both $P_0$ and $P_i$ . Consequently, Eve knows nothing about $K_i$ by launching this attack if not being detected.

(2) Trojan horse attack

The particles in $S_i$ ( $i=1,2,\ldots,N$ ) are transmitted forth and back between $P_0$ and $P_i$ . As a result, $P_i$ can places a wavelength filter and a photon number splitter (PNS) in front of her devices to avoid the invisible photon eavesdropping attack and the delay-photon Trojan horse attack, respectively [33,34].

(3) Intercept-resend attack

In order to know $K_i$ ( $i=1,2,\ldots,N$ ), Eve prepares the fake sequence $S_i^*$ in the $Z$ basis beforehand, intercepts $S_i$ from $P_0$ , and sends $S_i^*$ to $P_i$ ; after $P_i$ imposes her operations on $S_i^*$ , Eve intercepts the particle sequence sent out from $P_i$ and transmits $S_i$ to $P_0$ . Considering the case that $P_i$ chooses to REFLECT, $P_i$ reflects her $j$ th ( $j=1,2,\ldots,4n$ ) received fake particle to $P_0$ ; and $P_0$ measures the $j$ th particle of $S_i$ sent out from Eve and the corresponding particle in $T_i$ on her site with the $d$-dimensional Bell basis. Hence, Eve cannot be detected in this case. Considering the case that $P_i$ chooses to MEASURE, $P_i$ measures the $j$ th received fake particle with the $Z$ basis, prepares a fresh particle in the same state as that she found and sends it back to $P_0$ ; and $P_0$ measures the $j$ th particle of $S_i$ sent from Eve and the corresponding particle in $T_i$ on her site with the $Z$ basis. Hence, Eve can be discovered with the probability of $\frac{d-1}{d}$ in this case. It can be concluded now that when Eve launches this intercept-resend attack, she can be detected with the probability of $\frac{d-1}{2d}$ .

(4) Measure-resend attack

In order to know $K_i$ ( $i = 1,2,\ldots,N$ ), Eve intercepts $S_i$ from $P_0$ , adopts the $Z$ basis to measure its particles and transmits the corresponding resulted states to $P_i$ . Considering the case that $P_i$ chooses to REFLECT, $P_i$ reflects the $j$ th ( $j = 1,2,\ldots,4n$ ) received particle to $P_0$ ; and $P_0$ measures the $j$ th received particle from $P_i$ and the corresponding particle in $T_i$ on her site with the $d$-dimensional Bell basis. Consequently, Eve can be detected with the probability of $\frac{d-1}{d}$ in this case. Considering the case that $P_i$ chooses to MEASURE, $P_i$ measures the $j$ th received particle with the $Z$ basis, prepares a fresh particle in the found state and transmits it to $P_0$ . Apparently, in this case, Eve's attack cannot be detected. It can be concluded now that when Eve launches this measure-resend attack, the probability that she can be detected is $\frac{d-1}{2d}$ .

### 3.2 Participant attack

In this protocol, different $P_i$ s play the same role and are mutually independent, where $i = 1,2,\ldots,N$ . Two different cases of participant attack need to be discussed.

On one hand, suppose that only one of $P_1, P_2, \ldots, P_N$ is trustless. Without loss of generality, assume that $P_1$ is the trustless one. $P_1$ may impose her attacks on the transmitted particles from $P_0$ to $P_{\mathrm{m}}$ and back from $P_{\mathrm{m}}$ to $P_0$ , where $m = 2,3,\ldots,N$ . Unfortunately, just as analyzed above, because of being independent from $P_0$ and $P_{\mathrm{m}}$ , $P_1$ essentially acts as an outside eavesdropper and is undoubtedly discovered.

On the other hand, suppose that more than one of $P_1, P_2, \ldots, P_N$ is trustless. The worst situation is that $n-1$ parties of $P_1, P_2, \ldots, P_N$ are trustless. Without loss of generality, assume that $P_1$ is the only trustful one. $P_2, P_3, \ldots, P_N$ may launch their attacks on the transmitted particles from $P_0$ to $P_1$ and back from $P_1$ to $P_0$ . Unfortunately, due to being independent from $P_0$ and $P_1$ , $P_2, P_3, \ldots, P_N$ essentially act as an outside eavesdropper, which means that they are undoubtedly detected.

## 4 Discussions

In a quantum communication protocol within $d$-dimensional quantum system, we usually use the qudit efficiency to evaluate its performance of efficiency, which is defined as [35]

$$\eta = \frac{b}{q+c}. \tag{14}$$

Here, $b$ , $q$ and $c$ are the length of shared secret key, the number of qudits consumed and the number of classical information consumed, respectively. We do not consider the classical resources required for eavesdropping check processes.

In this protocol, the length of $K$ is $n$ , so it gets $b = n$ . $P_0$ needs to prepare $N$ $d$-dimensional Bell state sequences of length $4n$ and send $S_i$ to $P_i$ ; and when $P_i$ selects MEASURE for the received particles in $S_i$ , she needs to generate $2n$ new particles and transmit them to $P_0$ . As a result, it has $q = 4n \times 2 \times N + 2n \times N = 10nN$ . No classical resource is used during the classical communication, so $c = 0$ . Consequently, this protocol has the qudit efficiency of $\eta = \frac{n}{10nN} = \frac{1}{10N}$ .

We further compare this protocol with the ones of Ref.[32], which are the only two SQSS protocols feasible for $d$-dimensional quantum system up to now. It is easy to know from Table 1 that with respect to the number of initial quantum state, this protocol defeats the ones of Ref.[32]; and as for the usage of unitary operations, this protocol exceeds the ones of Ref.[32].

Table 1 Comparison among different MSQSS protocols

| | The first protocol of Ref.[32] | The second protocol of Ref.[32] | This protocol |
|---|---|---|---|
| Initial quantum resource | $d$ -dimensional single-particle states | $d$ -dimensional single-particle states | $d$ -dimensional Bell states |
| Number of initial quantum state | Multi-state | Multi-state | Single-state |
| Number of parties | Multiparty | Multiparty | Multiparty |
| Transmission mode | Tree-type | Circular | Tree-type |
| Usage of pre-shared key | No | No | No |
| Usage of quantum entanglement swapping | No | No | No |
| Usage of unitary operations | Yes | Yes | No |
| Usage of quantum measurement operation from the classical user | No | No | Yes |

## 5 Conclusions

A tree-type single-state MSQSS protocol with $d$-dimensional Bell states is designed in this paper. The security of this protocol against the outside attack and the participant attack is validated. This protocol is applicable for the $d$-dimensional quantum system, only employs one kind of $d$-dimensional Bell states as initial quantum resource, and needs neither quantum entanglement swapping nor unitary operations.

## Acknowledgments

Funding by the National Natural Science Foundation of China (Grant No.62071430 and No.61871347) and the Fundamental Research Funds for the Provincial Universities of Zhejiang (Grant No.JRK21002) is gratefully acknowledged.

## Reference

[1] Bennett, C.H., Brassard, G.: Quantum cryptography: public-key distribution and coin tossing. In: Proceedings of the IEEE International Conference on Computers, Systems and Signal Processing, pp.175-179. IEEE Press, Bangalore (1984)

[2] Ekert, A.K.: Quantum cryptography based on Bell's theorem. Phys. Rev. Lett. 67(6), 661 (1991)

[3] Long, G.L., Liu, X.S.: Theoretically efficient high-capacity quantum-key-distribution scheme. Phys. Rev. A 65, 032302 (2002)

[4] Bostrom, K., Felbinger, T.: Deterministic secure direct communication using entanglement. Phys. Rev. Lett. 89, 187902 (2002)

[5] Nguyen, B.A.: Quantum dialogue. Phys. Lett. A 328(1), 6-10 (2004)

[6] Ye, T.Y., Jiang L.Z.: Improvement of controlled bidirectional quantum direct communication using a GHZ state. Chin. Phys. Lett. 30(4), 040305 (2013)

[7] Ye, T.Y., Jiang, L.Z.: Quantum dialogue without information leakage based on the entanglement swapping between any two Bell states and the shared secret Bell state. Phys. Scr. 89(1), 015103 (2014)

[8] Ye, T.Y.: Robust quantum dialogue based on the entanglement swapping between any two logical Bell states and the shared auxiliary logical Bell state. Quantum Inf. Process. 14(4), 1469-1486 (2015)

[9] Ye, T.Y.: Quantum secure dialogue with quantum encryption. Commun. Theor. Phys. 62(3), 338-342 (2014)

[10] Ye, T.Y.: Fault tolerant channel-encrypting quantum dialogue against collective noise. Sci. China Phys. Mech. Astron. 58(4), 040301 (2015)

[11] Zhou, N.R., Zeng, G.H., Xiong, J.: Quantum key agreement protocol. Electron. Lett. 40, 1149-1150 (2004)

[12] Chong, S.K., Tsai, C.W., Hwang, T.: Improvement on "quantum key agreement protocol with maximally entangled states". Int. J. Theor. Phys. 50, 1793-1802 (2011)

[13] Hillery, M., Buzek, V., Berthiaume, A.: Quantum secret sharing. Phys. Rev. A 59, 1829 (1999)

[14] Karlsson, A., Koashi, M., Imoto, N.: Quantum entanglement for secret sharing and secret splitting. Phys. Rev. A 59(1), 162-168 (1999)

[15] Zhang, Z.J., Yang, J., Man, Z.X., Li, Y.: Multiparty secret sharing of quantum information using and identifying Bell state. Eur. Phys. J. D 33(1), 133-136 (2005)

[16] Deng, F.G., Zhou, H.Y., Long, G.L.: Circular quantum secret sharing. J. Phys. A: Gen. Phys. 39(45), 14089-14099 (2007)

[17] Deng, F.G., Li, X.H., Zhou, H.Y.: Efficient high-capacity quantum secret sharing with two-photon entanglement. Phys. Lett. A 372(12), 1957-1962 (2008)

[18] Sun, Y., Wen, Q.Y., Gao, F., Chen, X.B., Zhu, F.C.: Multiparty quantum secret sharing based on Bell measurement. Opt. Commun. 282(17), 3647-3651(2009)

[19] Gao, G., Wang, L.P.: An efficient multiparty quantum secret sharing protocol based on Bell states in the high dimension Hilbert space. Int. J. Theor. Phys. 49(11), 2852-2858 (2010)

[20] Hwang, T., Hwang, C.C., Li, C.M.: Multiparty quantum secret sharing based on GHZ states. Phys. Scr. 83, 045004 (2011)

[21] Boyer, M., Kenigsberg, D., Mor, T.: Quantum key distribution with classical Bob. Phys. Rev. Lett. 99(14), 140501 (2007)

[22] Ye, T.Y., Li, H.K., Hu, J.L.: Semi-quantum key distribution with single photons in both polarization and spatial-mode degrees of freedom. Int. J. Theor. Phys. 59, 2807-2815 (2020)

[23] Ye, T.Y., Geng, M.J., Xu, T.J., Chen, Y.: Efficient semiquantum key distribution based on single photons in both polarization and spatial-mode degrees of freedom. Quantum Inf. Process. 21, 123 (2022)

[24] Li, Q., Chan, W.H., Long, D.Y.: Semiquantum secret sharing using entangled states. Phys. Rev. A 82, 022303 (2010)

[25] Wang, J., Zhang, S., Zhang, Q., Tang, C.J.: Semiquantum key distribution using entangled states. Chin. Phys. Lett. 28(10), 100301 (2011)

[26] Li, L., Qiu, D., Mateus, P.: Quantum secret sharing with classical Bobs. J. Phys. A Math. Theor. 46(4), 045304 (2013)

[27] Yang, C.W., Hwang, T.: Efficient key construction on semi-quantum secret sharing protocols. Int. J. Quantum Inf. 11(5), 1350052 (2013)

[28] Gao, G., Wang, Y., Wang, D.: Multiparty semiquantum secret sharing based on rearranging orders of qubits. Mod. Phys. Lett. B 30 (10), 1650130(2016)

[29] Ye, C.Q., Ye, T.Y.: Circular semi-quantum secret sharing using single particles. Commun. Theor. Phys. 70, 661-671 (2018)

[30] Tsai, C.W., Yang, C.W., Lee, N.Y.: Semi-quantum secret sharing protocol using W-state. Mod. Phys. Lett. A 34(27), 1950213 (2019)

[31] Li, C.Y, Ye, C.Q., Tian, Y., Chen, X.B., Li, J.: Cluster-state-based quantum secret sharing for users with different abilities. Quantum Inf. Process. 20(12), 1-14 (2021)

[32] Ye, C.Q., Ye, T.Y., He, D., et al.: Multiparty semi-quantum secret sharing with $d$ -level single-particle states. Int. J. Theor. Phys. 58(11), 3797-3814 (2019)

[33] Deng, F.G., Zhou, P., Li, X.H., et al.: Robustness of two-way quantum communication protocols against Trojan horse attack. https://arxiv.org/abs/quant-ph/0508168 (2005)

[34] Li, X.H., Deng, F.G., Zhou, H.Y.: Improving the security of secure direct communication based on the secret transmitting order of particles. Phys. Rev. A 74, 054302 (2006)

[35] Geng, M.J., Xu, T.J., Chen, Y., Ye, T.Y.: Semiquantum private comparison of size relationship based $d$ -level single-particle states. Sci. Sin. Phys. Mech. Astron. 52(9), 290311 (2022)